\documentclass[twocolumn,showpacs,preprintnumbers,amsmath,amssymb]{revtex4}

\usepackage{graphicx}
\usepackage{dcolumn}
\usepackage{bm}

\begin{document}

\title{Ghost imaging in scattering media}

\author{Wenlin Gong}
\author{Pengli Zhang}
\author{Xia Shen}
\author{ Shensheng Han}
\email{sshan@mail.shcnc.ac.cn} \affiliation{ Key Laboratory for
Quantum Optics and Center for Cold Atom Physics, Shanghai Institute
of Optics and Fine Mechanics, Chinese Academy of Sciences, Shanghai
201800, China }

\date{\today}

\begin{abstract}
Ghost imaging with thermal light in scattering media is
investigated. We demonstrated both theoretically and experimentally
for the first time that the image with high quality can still be
obtained in the scattering media by ghost imaging. The scattering
effect on the qualities of the images obtained when the object is
illuminated directly by the thermal light and ghost imaging is
analyzed theoretically. Its potential applications are also
discussed.
\end{abstract}


\maketitle
\section{introduction}
Multiple scattering has a great effect on the qualities of images
and the transmission of information. The information will be decayed
and the images suffer reduced resolution and contrast because of
multiple scattering. For example, the measurement of the laser radar
\cite{Kamerman}, satellite communications \cite{Booker}, the
propagation and imaging of light in the atmosphere \cite{Mckechnie},
neutron imaging \cite{Raine} and the imaging and diagnosis in life
and medical science \cite{Maher}. So the imaging in strong
scattering media is always a great problem and presents a key
challenge for the research of better imaging method and technique.

In clinic applications, the most common imaging modalities include
ultrasound imaging, X-ray computed tomography(CT), and magnetic
resonance imaging (MRI) \cite{Maher,Wang}. As the development of
imaging technology, optical imaging is becoming an increasing
interesting method for the imaging in biological tissue. By now most
imaging methods are obtained by using the gating techniques. Such as
confocal imaging, spatial filtering, optical coherence tomography
(OCT), Mueller optical coherence tomography, Diffuse optical
tomography (DOT), Photoacoustic tomography (PAT),
Ultrasound-Modulated optical tomograph (UOT) and so on
\cite{Wang,Gibson,Yodh,Arridge1,Arridge2,Sachin,Hebden,Chance,Laubscher,Brian}.
Although the qualities of the images have a great increase by these
techniques, there is still lots of problems which are difficult to
be done. Because the imaging techniques in scattering media
discussed above mainly are only the first-order effect of light
field, detection and imaging are unseparated. When the information
of the object is distorted by multiple scattering, and the
information of both multiple scattering and the object is unknown,
so we can not, in principle, obtain exactly the images destroyed by
the multiple scattering, which leads to be impossible of the
restoration of the qualities of images caused by multiple
scattering.

The first two-photon imaging experiment with entangled source was
demonstrated by Pittman \emph{etal}. in 1995 \cite{Pittman}, which
shew that we could obtain a nonlocal image by transmitting pairs of
photons through a test and a reference path. Since 2002, the
theories and experiments demonstrated that the ghost imaging could
also be obtained with thermal light
\cite{Cheng,Gatti1,Gatti2,Valencia,Bennink,Basano}. And the fierce
discussion on the essence of ghost imaging at one time
\cite{Scarcelli1,Gatti3,Scarcelli2}. Ghost imaging is considered as
the effect of second-order correlation of light field and is caused
by the undistinguishable relation of identical particles
\cite{Zhang1,Scarcelli1,Scarcelli2}.  For the first time, detection
and imaging are separated by ghost imaging. The test path and the
reference path are used to detect the information from the object
and imaging for the object, respectively. Recently we find that the
qualities of ghost images are determined by both the reference path
and test path \cite{Zhang}. Because multiple scattering only
degrades the imaging quality in the test path, whereas there is no
multiple scattering in the reference path. By correlation
measurement, we may get a image with much better quality than the
image obtained by detection in a single path.

\section{theory and analysis}
\begin{figure}
\centerline{
\includegraphics[width=8cm]{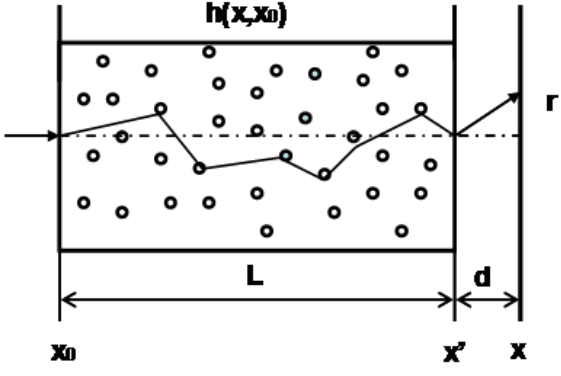}}
\caption{Schematic diagram for the light transmitting in a
scattering media.}
\end{figure}

Schematic diagram for the light transmitting in a scattering media
is shown in Fig. 1. In the theory of linear systems \cite{Goodman},
the light field $E(x)$ on the plane $x$ is the convolution of the
light field $E(x_0)$ on the plane $x_0$ and the impulse response
function $h(x,x_0)$.

\begin{eqnarray}
E(x) = \int {dx_0 E(x_0 )} h(x,x_0 ).
\end{eqnarray}

For light transmission in scattering media, the light field on the
position $x$ is the linear superposition the incident light and the
scattering light.
\begin{eqnarray}
E(x) = \alpha \int {dx_0 } E(x_0 )h_{in} (x,x_0 ) \nonumber\\+ \beta
\int d x_0 E(x_0 )h_{sca} (x,x_0 ),
\end{eqnarray}
\begin{eqnarray}
\left| \alpha  \right|^2  + \left| \beta  \right|^2  = 1.
\end{eqnarray}
where $h_{in}(x,x_0)$ is the impulse response function with no
scattering media, and $h_{sca}(x,x_0)$ is the impulse response
function from the plane $x_0$ to the plane $x$ because of the
interactions of multiple scattering, and $\alpha$, $\beta$ are the
probability amplitudes of the incident light and the scattering
light, respectively. From Eqs. (1)-(3), we have

\begin{eqnarray}
h(x,x_0 ) = \alpha h_{in} (x,x_0 ) + \beta h_{sca} (x,x_0 ).
\end{eqnarray}

The probability distribution function in scattering media is called
point scattering function. The impulse response function
$h_{sca}(x,x_0)$ has close contact with the point scattering
function which is Dirac delta function when there is no scattering
media. However, in the scattering media, it is a spread function
with a broadening length, and generally the point scattering
function has two forms: Lorentzian-shaped and Gaussian-shaped
distribution \cite{Hassanein,Segre}. In multiple scattering Mie
theory\cite{Mie,Toublanc,Sharma}, both of probability amplitudes
$\alpha$, $\beta$ are depending on the diameter size of the particle
$D$, the wavelength of the incident light $\lambda$, the
concentration of suspended particles $w$ and the effective length of
scattering media $L$. According to the experiments and theories
\cite{Jessica,Milun,Mustafa,Faulkner,wells,yura,Hassanein,Segre}, we
have
\begin{subequations}
\label{eq:whole}
\begin{eqnarray}
\alpha  = \alpha (D,\lambda ,w,L) \propto \frac{{\lambda ^{b_\alpha
} }}{{D^{a_\alpha  } w^{c_\alpha  } L^{d_\alpha  } }},
\end{eqnarray}
\begin{eqnarray}
\beta  = \beta (D,\lambda ,w,L) \propto \frac{{D^{a_\beta  }
w^{c_\beta  } L^{d_\beta  } }}{{\lambda ^{b_\beta  } }},
\end{eqnarray}
\begin{eqnarray}
h_{sca} (x,x_0 ) \propto \int dx' P(x',x_0 )_{L_A} h(x,x')_{(L+d)},
\end{eqnarray}
\begin{eqnarray}
P(x',x_0 )_{L_A} =[\frac{2}{{\pi \Delta x_{L_A } ^2 }}]^{1/4}
\nonumber\\\times\exp \left\{ { - (\frac{{x' - x_0 }}{{\Delta
x_{L_A} }})^2 } \right\},
\end{eqnarray}
\begin{eqnarray}
\Delta x_{L_A }\propto \frac{{D^{a_x  } w^{c_x  } {L_A }^{d_x }
}}{{\lambda ^{b_x  } }},
\end{eqnarray}
\begin{eqnarray}
\int {\left| {P(x',x_0 )_{L_A} } \right|} ^2 dx'=1.
\end{eqnarray}
\end{subequations}
where $P(x',x_0 )_{L_A}$ is point scattering probability amplitude.
$\Delta x_{L_A}$ is broadening length because of the interactions of
multiple scattering, and it becomes wider with the increase of the
scattering length. With the increase of the broadening length, the
frequency spectrum of the optical transfer function becomes
narrower, which is the main reason leading to the degradation of the
quality of information transmission and images \cite{wells,yura}. We
suppose point scattering function is Gaussian-shaped distribution
without considering the absorption of scattering media. All the
coefficients in Eq. (5) should be determined by specific
experimental conditions.

The scheme for ghost imaging with thermal light in the scattering
media is shown in Fig. 2. The light source $S$, first propagates
through a beam splitter, then is divided into a test and a reference
path. In the test path, the light propagates through a single lens
of focal length $f_1$, the scattering media and then to the detector
$D_t$. In the reference path, the light propagates through a single
lens of focal length $f$ then to an array of pixel detector $D_r$.

\begin{figure}
\centerline{
\includegraphics[width=8cm]{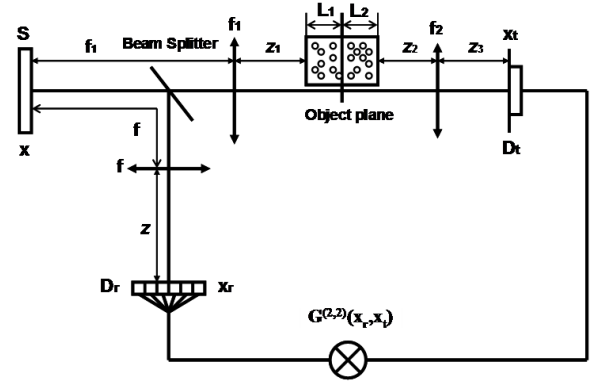}}
\caption{Scheme for ghost imaging with thermal light in the
scattering media.}
\end{figure}

By optical coherence theory \cite{Cheng,Glauber}, we can obtain the
correlation function of intensity fluctuations between the
detectors:

\begin{subequations}
\label{eq:whole}
\begin{eqnarray}
\Delta G^{(2,2)} (x_r ,x_t ) = \left\langle {\Delta I_r (x_r )\Delta
I_t (x_t )} \right\rangle  = \left| {\Gamma (x_r ,x_t )} \right|^2,
\end{eqnarray}
\begin{eqnarray}
\Gamma (x_r ,x_t )= \int {dx_1 } \int {dx_2 G^{(1,1)} (x_1 ,x_2
)h_r^*(x_r,x_1)}\nonumber\\\times h_t(x_t ,x_2).
\end{eqnarray}
\end{subequations}
where $\Gamma(x_r ,x_t )$ is the first-order cross-correlation
function of two different points from the test and reference paths.

Suppose the light source is fully spatially incoherent, then
\begin{equation}
G^{(1,1)} (x_1 ,x_2 ) = I_0 \delta (x_1  - x_2 ).
\end{equation}
where $I_0$ is a constant, and $\delta(x)$ is the Dirac delta
function.

Under the paraxial and small angle approximation, and when the
effective apertures of the lenses in the optical system are large
enough, the impulse response function of the reference system is
\begin{equation}
h_r (x_r ,x_1) \propto \exp \left\{ {\frac{{j\pi }}{{\lambda f}}(1 -
\frac{z}{f})x_1 ^2  - \frac{{2j\pi }}{{\lambda f}}x_r x_1 }
\right\}.
\end{equation}
And $f_2,z_2+L_2,z_3$ obey the Lens Law
\begin{equation}
\frac{1}{{z_2+L_2 }} + \frac{1}{{z_3 }} = \frac{1}{{f_2 }}.
\end{equation}
when
\begin{eqnarray}
\frac{{f_1 }}{f} = \frac{{z_2  + L_2 }}{{z_3 }}.
\end{eqnarray}
The impulse response function of the test system is

\begin{subequations}
\label{eq:whole}
\begin{eqnarray}
h(x_t ,x_2 ) \propto \int {dx'} [\alpha _1 \exp \left\{ { -
\frac{{2j\pi }}{{\lambda f_1 }}x'x_2 } \right\} + \beta _1 \int d
x_2 '\nonumber\\\times P(x',x_2 ')_{L_{1A} } \exp \left\{ {-
\frac{{2j\pi }}{{\lambda f_1 }}x_2 'x_2 } \right\}]
t(x')C(x')\nonumber\\\times\exp \left\{ {\frac{{j\pi }}{{\lambda f_1
}}(1 - \frac{{z_1 + L_1 }}{{f_1 }})x_2 ^2 } \right\},
\end{eqnarray}

\begin{eqnarray}
C(x) = [\alpha _2 \delta (x + \frac{{f_1 }}{f}x_t )\exp \left\{
{\frac{{j\pi }}{{\lambda (L_2+z_2) }}x^2 } \right\} + \beta _2
\nonumber\\\times P( - \frac{{f_1 }}{f}x_t ,x)_{L_{2A} } \exp
\left\{ {\frac{{j\pi }}{{\lambda (L_2+z_2) }}(\frac{{f_1 }}{f}x_t)^2
} \right\}].
\end{eqnarray}
\end{subequations}

Substituting Eqs. (7) and (9)-(11b) into Eq. (6b), we can get the
intensity distribution in the test path

\begin{eqnarray}
I(x_t ) \propto \int {dx'} \int {dx''}t(x')t^ *  (x'') \{ \left|
{\alpha _1 } \right|^2 \delta (x' - x'') \nonumber\\+
2P(x',x'')_{L_{1A} } {\mathop{\rm Re}\nolimits} [\alpha _1 ^ * \beta
_1 ]  + \left| {\beta _1 } \right|^2 \int {dx_2 '} \nonumber\\\times
P(x',x_2 ')_{L_{1A} } P(x'',x_2 ')_{L_{1A} } \}C(x')C^ * (x'').
\end{eqnarray}
Eq. (12) describes the images in the scattering media for the direct
illumination of the thermal light . When
\begin{eqnarray}
\frac{{1 - \frac{z}{f}}}{f} = \frac{{1 - \frac{{z_1  + L_1 }}{{f_1
}}}}{{f_1 }}.
\end{eqnarray}
substituting Eqs. (7)-(11b) and (13) into Eq. (6), we can obtain the
correlation function of intensity fluctuations

\begin{subequations}
\label{eq:whole}
\begin{eqnarray}
\Delta G^{(2)} (x_r, x_t ) \propto \left| {\alpha _1 \alpha _2 C_1
+ \alpha _1 \beta _2 C_2 } \right.\nonumber\\\left. { + \beta _1
\alpha _2 C_3  + \beta _1 \beta _2 C_4 } \right|^2,
\end{eqnarray}
\begin{eqnarray}
C_1  = \delta (x_r  + x_t )t( - \frac{{f_1 }}{f}x_t ),
\end{eqnarray}
\begin{eqnarray}
C_2  = t(\frac{{f_1 }}{f}x_r )P( - \frac{{f_1 }}{f}x_t ,\frac{{f_1
}}{f}x_r )_{L_{2A} },
\end{eqnarray}
\begin{eqnarray}
C_3  = t( - \frac{{f_1 }}{f}x_t )P( - \frac{{f_1 }}{f}x_t
,\frac{{f_1 }}{f}x_r )_{L_{1A} },
\end{eqnarray}
\begin{eqnarray}
C_4  = \int {dx't(x')} P(x',\frac{{f_1 }}{f}x_r )_{L_{1A} } P( -
\frac{{f_1 }}{f}x_t ,x')_{L_{2A} }.
\end{eqnarray}
\end{subequations}

The ghost imaging in the scattering media is described by the Eqs.
(14a)-(14e). And it is a image with high quality for $C_1$ and
$C_2$, which implies we can still get a image with high quality by
ghost imaging in the scattering media.

From Eqs. (14a)-(14e), if the test detector is an array of pixel
detector, and $x_t=-x_r$, after some calculation, then
\begin{eqnarray}
\Delta G^{(2)} (x_r , - x_r ) \propto \left| {\left\{ {\alpha _1
\alpha _2  + \alpha _1 \beta _2 \left[ {\frac{2}{{\pi \Delta
x_{L_{2A} } ^2 }}} \right]^{1/4} } \right.} \right.\nonumber\\\left.
{ + \beta _1 \alpha _2 \left. {\left[ {\frac{2}{{\pi \Delta
x_{L_{1A} } ^2 }}} \right]^{1/4} } \right\}t(\frac{{f_1 }}{f}x_r ) +
\beta _1 \beta _2 C_4 } \right|^2 .
\end{eqnarray}
if the test detector is a bucket detector, then
\begin{eqnarray}
\Delta G^{(2)} (x_r) \propto \int {dx_t}\left| {\alpha _1 \alpha _2
C_1 + \alpha _1 \beta _2 C_2 } \right.\nonumber\\\left. { + \beta _1
\alpha _2 C_3  + \beta _1 \beta _2 C_4 } \right|^2.
\end{eqnarray}
Eqs. (15) and (16) represent ghost imaging when the test detector is
an array of pixel detector or a bucket detector, respectively. From
Eqs. (15), if $\beta_1=0$(namely $ L_1=0$) or $\beta_2=0$(namely
$L_2=0$), it is clear to find that we can still obtain a image with
high quality by ghost imaging in the scattering media. if
$\beta_1\neq0$ and $\beta_2\neq0$ and as the increase of $\beta_1$
and $\beta_2$, the qualities of ghost images will reduce, and the
term including $C_4$ is the main reason leading to the degradation
of the qualities of the images.

For $L_1=0$, there is only multiple scattering between the object
plane and the test detector, then we have $\left| \alpha_1 \right| =
1,\beta _1 = 0$. By Eq. (12), the intensity distribution in the test
path is
\begin{subequations}
\label{eq:whole}
\begin{eqnarray}
I_t (x_t ) \propto (\left| {\alpha _2 } \right|^2  + 2C_{5t} )\left|
{t( - \frac{{f_1 }}{f}x_t )} \right|^2  + C_{6t} \left| {\beta _2 }
\right|^2 ,
\end{eqnarray}
\begin{eqnarray}
C_{5t}  = [\frac{2}{{\pi \Delta x_{L_{2A} } ^2 }}]^{1/4}
{\mathop{\rm Re}\nolimits} [\alpha _2 ^ * \beta _2 ],
\end{eqnarray}
\begin{eqnarray}
C_{6t}  = \int {dx'} \left| {t(x')} \right|^2 \left| {P ( -
\frac{{f_1 }}{f}x_t ,x')_{L_{2A} } } \right|^2\nonumber\\ =
[\frac{2}{{\pi \Delta x_{L_{2A} } ^2 }}]^{1/2}\int {dx'} \left|
{t(x')} \right|^2 \nonumber\\\times\exp \left\{ { - \frac{2}{{\Delta
x_{L_{2A}} ^2 }}(x' + \frac{{f_1 }}{f}x_t )^2 } \right\} .
\end{eqnarray}
\end{subequations}
the last term in Eq. (17a) is the main reason leading to the
decrease of the quality of the image when there is multiple
scattering between the object plane and the detector. With the
increase of $\beta_2$ (and the decrease of the probability amplitude
$\alpha_2$), the quality of the image will be further degraded.

Form Eq. (15), when the test detector is an array of pixel detector,
after some calculation, then
\begin{eqnarray}
\Delta G^{(2)} (x_r , - x_r ) \propto \left| {\alpha _2  + \beta _2
\left[ {\frac{2}{{\pi \Delta x_{L_{2A} } ^2 }}} \right]^{1/4} }
\right|^2 \nonumber\\\times\left| {t(\frac{{f_1 }}{f}x_r )}
\right|^2.
\end{eqnarray}

If the test detector is a bucket detector, By Eq. (16), then
\begin{subequations}
\label{eq:whole}
\begin{eqnarray}
\Delta G^{(2)} (x_r) \propto (\left| \alpha_2  \right|^2  + 2C_5
+C_6 \left| \beta_2 \right|^2)\left| {t(\frac{{f_1 }}{f}x_r )}
\right|^2,
\end{eqnarray}
\begin{eqnarray}
C_5  =[\frac{2}{{\pi \Delta x_{L_{2A} } ^2 }}]^{1/4}{\mathop{\rm
Re}\nolimits} [\alpha _2 ^ *  \beta _2 ],
\end{eqnarray}
\begin{eqnarray}
C_6  = \int {dx_t \left| {P( - \frac{{f_1 }}{f}x_t ,\frac{{f_1
}}{f}x_r )_{L_{2A} } } \right|^2 }  \sim 1.
\end{eqnarray}
\end{subequations}
from Eqs. (18)-(19c), we find that whether the test detector is an
array of pixel detector or a bucket detector, the qualities of ghost
images can be obtained even though there is multiple scattering
between the object plane and the test detector.

For $L_2=0$, there is only multiple scattering between the source
and the object plane, then we can gain $\left| \alpha_2 \right| =
1,\beta _2 = 0$. By Eq. (12), the intensity distribution in the test
path is
\begin{subequations}
\label{eq:whole}
\begin{eqnarray}
I_t (x_t ) \propto (\left| {\alpha _1 } \right|^2  + 2C_{7t}  +
C_{8t} \left| {\beta _1 } \right|^2 )\left| {t( - \frac{{f_1
}}{f}x_t )} \right|^2,
\end{eqnarray}
\begin{eqnarray}
C_{7t}  = [\frac{2}{{\pi \Delta x_{L_{1A} } ^2 }}]^{1/4}
{\mathop{\rm Re}\nolimits} [\alpha _1 ^ * \beta _1 ],
\end{eqnarray}
\begin{eqnarray}
C_{8t}  = \int {dx_2 '} \left| {P( - \frac{{f_1 }}{f}x_t ,x_2
')_{L_{1A} } } \right|^2  \sim 1.
\end{eqnarray}
\end{subequations}
From Eqs. (20a)-(20c), we find that the multiple scattering between
the source and the object plane has no effect on the quality of the
image.

From Eq. (15), when the test detector is an array of pixel detector,
after some calculation, then
\begin{eqnarray}
\Delta G^{(2)} (x_r , - x_r ) \propto \left| {\alpha _1  + \beta _1
\left[ {\frac{2}{{\pi \Delta x_{L_{1A} } ^2 }}} \right]^{1/4} }
\right|^2 \nonumber\\\times\left| {t(\frac{{f_1 }}{f}x_r )}
\right|^2.
\end{eqnarray}
which reveals that we can still obtain a image with high quality
when the test detector is an array of pixel detector even if there
is multiple scattering between the source and the object plane.

If the test detector is a bucket detector, By Eq. (16), then
\begin{subequations}
\label{eq:whole}
\begin{eqnarray}
\Delta G^{(2)} (x_r ) \propto (\left| \alpha_1  \right|^2  + 2C_7
)\left| {t(\frac{{f_1 }}{f}x_r )} \right|^2  + C_8 \left| \beta_1
\right|^2,
\end{eqnarray}
\begin{eqnarray}
C_{7t}  = [\frac{2}{{\pi \Delta x_{L_{1A} } ^2 }}]^{1/4}
{\mathop{\rm Re}\nolimits} [\alpha _1 ^ * \beta _1 ],
\end{eqnarray}
\begin{eqnarray}
C_8  = \int {dx_t } \left| {t( - \frac{{f_1 }}{f}x_t )P( -
\frac{{f_1 }}{f}x_t ,\frac{{f_1 }}{f}x_r )_{L_{1A} } } \right|^2
\nonumber\\= \left[ {\frac{2}{{\pi \Delta x_{L_{1A} } ^2 }}}
\right]^{1/2} \int {dx_t } \left| {t( - \frac{{f_1 }}{f}x_t )}
\right|^2 \nonumber\\\times\exp \left\{ { - \frac{{2f_1 ^2
}}{{\Delta x_{L_{1A} } ^2 f^2 }}(x_t  + x_r )^2 } \right\}.
\end{eqnarray}
\end{subequations}
where $C_8$ is the main factor leading to the decrease of the
qualities of ghost images. $f_1>f$ is helpful to improve the quality
of the image.

The second-order degree of coherence $g^{(2)}(x_1,x_2)$ at the
positions $x_1$ and $x_2$ can be defined as follows
\begin{eqnarray}
g^{(2)}(x_1,x_2)  = \frac{{\left\langle {I_1  \cdot I_2 }
\right\rangle }}{{\left\langle {I_1 } \right\rangle \left\langle
{I_2 } \right\rangle }} = 1 + \frac{{\left\langle {\Delta I_1  \cdot
\Delta I_2 } \right\rangle }}{{\left\langle {I_1 } \right\rangle
\left\langle {I_2 } \right\rangle }}.
\end{eqnarray}
for ghost imaging, The second-order degree of coherence
$g^{(2)}(x_{t1},x_{t2})$ obtained only by the correlation
measurement in the test path reveals the characteristic of the
source, whereas $g^{(2)}(x_r,x_t)$ describes the correlation between
two paths. And the degradation of $g^{(2)}(x_r,x_t)$ will lead to
the decay of intensity fluctuations, which has a great effect on the
visibility of ghost images.

Figs. 3-7 present numerical results of imaging a single slit in
scattering media based on Eqs. (12) and (15)-(16) (in which we take
$\lambda$=650nm, $f_1$=400mm, $f$=250mm, the single slit width
a=0.2mm). From Fig. 3(a), as the increase of $L_2$, the qualities of
the images will decrease obviously when the object is illuminated
directly by the thermal light. The quality of the image also degrade
rapidly with the increase of the broadening length $\triangle x_2$
and the decrease of the probability amplitude $\alpha_2$ when the
object is fixed in the position of $L_1$=0mm, $L_2$=100mm (Fig.
3(b), (c)), which accord with the results described by the Eqs
(17a)-(17c).

\begin{figure}
\centerline{
\includegraphics[width=8cm]{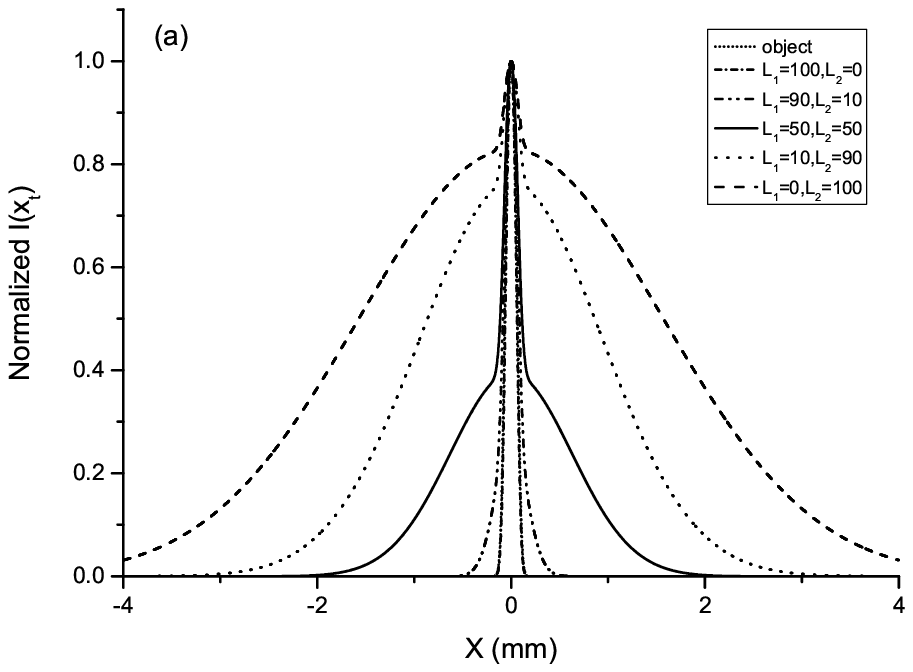}}
\end{figure}
\begin{figure}
\centerline{
\includegraphics[width=8cm]{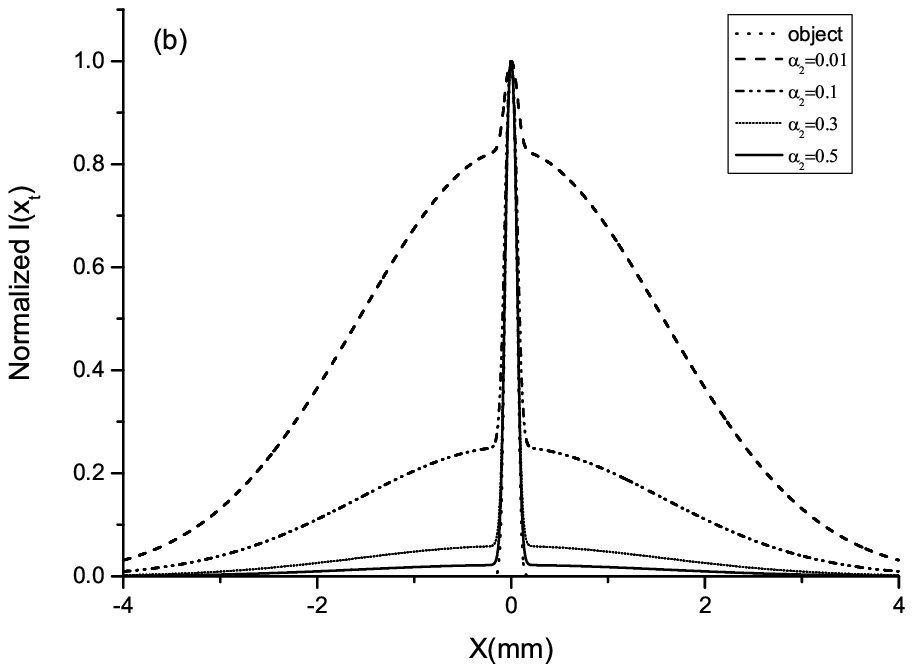}}
\end{figure}
\begin{figure}
\centerline{
\includegraphics[width=8cm]{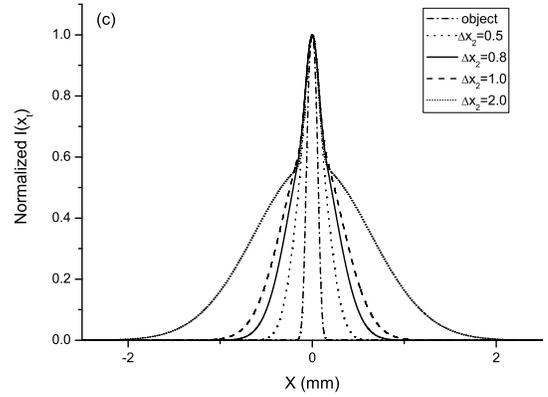}}
\caption{Factors which have the effect on the qualities of the
images when the object is illuminated directly by the thermal light.
(a). Images of a single slit at different positions in the
scattering media; (b). Images of a single slit for different
probability amplitude $\alpha_2$ when $L_1$=0mm, $L_2$=100mm and the
broadening length $\triangle x_2$=5.0; and (c). Images of a single
slit for different probability amplitude $\triangle x_2$ when
$L_1$=0mm, $L_2$=100mm and the broadening length $\alpha_2$=0.05.}
\end{figure}

Fig. 4 shows the numerical results when the position of the object
in the scattering media is shifted. By Fig. 4(a), we can find that
if the test detector is an array of pixel detector, multiple
scattering has no effect on the qualities of ghost images when there
is only multiple scattering either between the object plane and the
test detector or between the object plane and the source. As the
object is fixed closing to the middle of scattering media, the
qualities of ghost images will decay. However, if the test detector
is a bucket detector, the qualities of ghost images will decrease
sharply with the degradation of $L_1$, which is opposite absolutely
to imaging when the object is illuminated directly by the thermal
light (Fig. 3(a)).

\begin{figure}
\centerline{
\includegraphics[width=8cm]{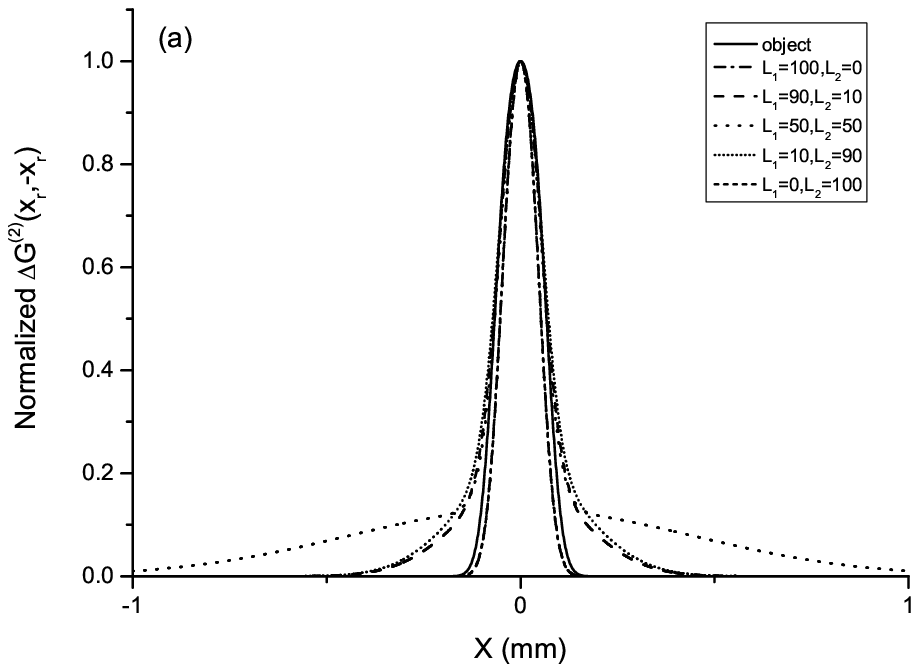}}
\end{figure}
\begin{figure}
\centerline{
\includegraphics[width=8cm]{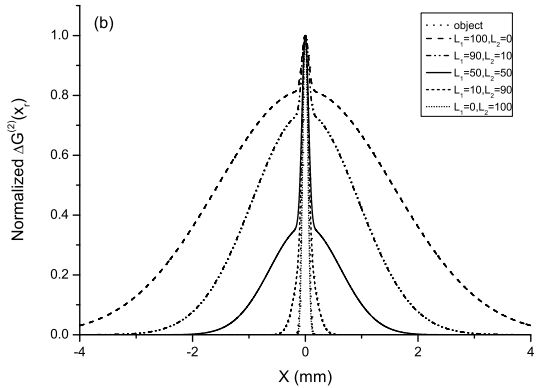}}
\caption{Relationship between the qualities of ghost images and the
position of the object in scattering media. (a). the test detector
is an array of pixel detector; and (b). the test detector is a
bucket detector.}
\end{figure}

In Fig. 5, We give the comparison of the qualities of images between
imaging by the direct illumination of the thermal light and ghost
imaging when the object is fixed in the middle of scattering media.
It is easy to find that we can only obtain the same image with low
quality as imaging by the direct illumination of the thermal light
when the test detector is a bucket detector by ghost imaging. But a
image with high quality can still be gained if the test detector is
an array of pixel detector.

\begin{figure}
\centerline{
\includegraphics[width=8cm]{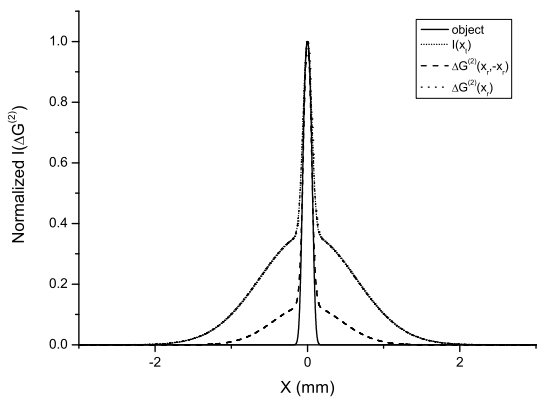}}
\caption{Comparison between imaging by the direct illumination of
the thermal light and ghost imaging when $L_1$=50mm, $L_2$=50mm, the
broadening length $\triangle x_1$=2.0, $\triangle x_2$=2.0 and
$\alpha_1$=0.1, $\alpha_2$=0.1.}
\end{figure}

Results shown in Fig. 6 reveals that the qualities of ghost images
will reduce obviously as the degradation of the probability
amplitude $\alpha_1$ and $\alpha_2$. And the object is still fixed
in the middle of scattering media.

\begin{figure}
\centerline{
\includegraphics[width=8cm]{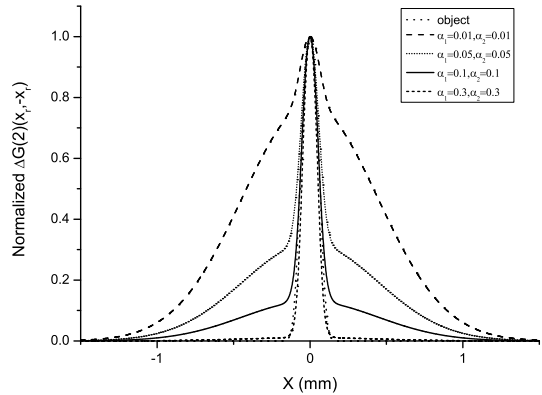}}
\caption{Dependence of probability amplitude $\alpha_2$ on the
qualities of ghost images when $L_1$=50mm, $L_2$=50mm and the
broadening length $\triangle x_1$=2.0, $\triangle x_2$=2.0.}
\end{figure}

In order to improve the qualities of ghost images further, we
investigate a new method to the effect on the qualities of ghost
images by using a circular aperture gating to change the transverse
coherent length near the scattering media in Fig. 7, which is
different from the method to change the transverse size of the
source. But when the space interval $\Delta s$ is greater than the
characteristic scale of the object, the diffraction will emerge.

\begin{figure}
\centerline{
\includegraphics[width=8cm]{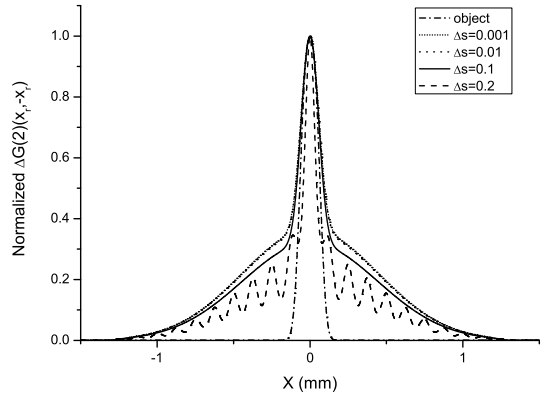}}
\caption{The effect of the space interval $\Delta s$ on the
qualities of ghost images when a circular aperture gating with
different space interval $\Delta s$ is fixed on the plane $x_0$.
$L_1$=50mm, $L_2$=50mm, the broadening length $\triangle x_1$=2.0,
$\triangle x_2$=2.0 and $\alpha_1$=0.1, $\alpha_2$=0.1.}
\end{figure}

\section{experiment}
In the experiments, we prepare a suspension liquid which is composed
by emulsion polymerization particles with particle diameter
$D$=3.26$\mu$m and the solution $NaCl$ with the density $\rho$=1.19
$g/cm^3$. The vessel used to put the suspension liquid is designed
as 40mm$\times$40mm$\times$20mm. The liquid can be considered as
strong multiple scattering media. And we take $\lambda$=650nm,
$f_1$=400mm, $f$=250mm, $z$=211mm, $z_1$=300mm, $z_2$=390mm,
$z_3$=243.8mm. The minimum characteristic scale of the object
($\textbf{`zhong'}$ ring) is 60 $\mu$m and the diameter of the ring
is 1.6mm. And the detectors in both paths are arrays of pixel
detectors.

Images shown in Fig. 8 (1) and (2) were the experimental results of
the object (\textbf{`zhong'} ring) by direct illumination and ghost
imaging when the object was fixed at different positions in the
scattering media, respectively. Form Fig. 8, when the object is
illuminated directly by the thermal light, the qualities of images
will reduce as the increase of the length $L_2$ of scattering media.
However, when there is only strong multiple scattering between the
object plane and the source or between the object plane and the
detector $D_t$, we can both obtain ghost images with high qualities.
If the object is fixed in the middle of the scattering media, the
visibility of ghost images will reduce, but the resolution doesn't
degrade. All discussed above accord with the theoretical results
described by the Eqs. (1).

\begin{figure}
\centerline{
\includegraphics[width=8cm]{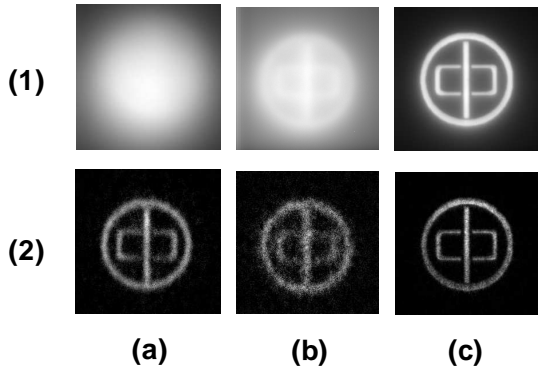}}
\caption{Images of the aperture (\textbf{`zhong'} ring) when the
object is fixed in different position of the scattering media. (a).
$L_1$=0mm, $L_2$=40mm; (b). $L_1$=20mm, $L_2$=20mm; and (c).
$L_1$=40mm, $L_2$=0mm. (1). when the object was illuminated directly
by thermal light; (2). ghost imaging.}
\end{figure}

In Fig. 9, we demonstrated experimentally that the effect of the
concentration of scattering media or the coherent length located at
the object plane on ghost imaging when the object is fixed in the
middle of scattering media. The visibility of ghost images will
reduce with the increase of the concentration of scattering media.
When the coherent length located at the object plane becomes long,
the visibility of ghost images will improve. However, the resolution
will degrade.

\begin{figure}
\centerline{
\includegraphics[width=8cm]{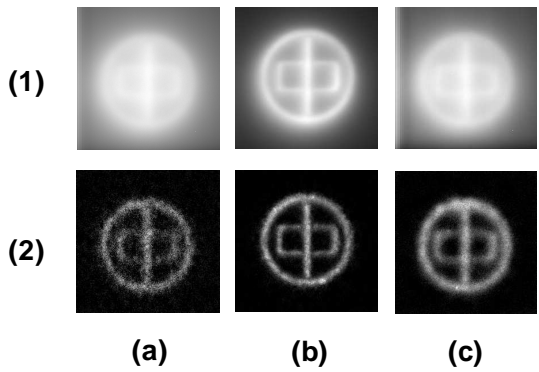}}
\caption{Effect of the concentration of scattering media and the
coherent length located at the object plane on ghost imaging. (a).
$L_c$=40.6$\mu$m, 6 drops; (b). $L_c$=40.6$\mu$m, 3 drops; and (c).
$L_c$=192.5$\mu$m, 6 drops. (1). when the object was illuminated
directly by thermal light; (2). ghost imaging.}
\end{figure}

Carves shown in Figs. 10 and 11 are the effect of concentration of
emulsion polymerization particles on $g^{(2)}$ when $L_1$=40mm and
$L_2$=0mm and when $L_1$=20mm and $L_2$=20mm without the object,
respectively. We can find that $g^{(2)}(x_{t1},x_{t2})$ and
$g^{(2)}(x_r,x_t)$ decay sharply with the increase of concentration
of emulsion polymerization particles. When there is two drops
emulsion polymerization particles put into the solution $NaCl$, the
cross-correlation coefficient will be lower than 1.10.

\begin{figure}
\centerline{
\includegraphics[width=8cm]{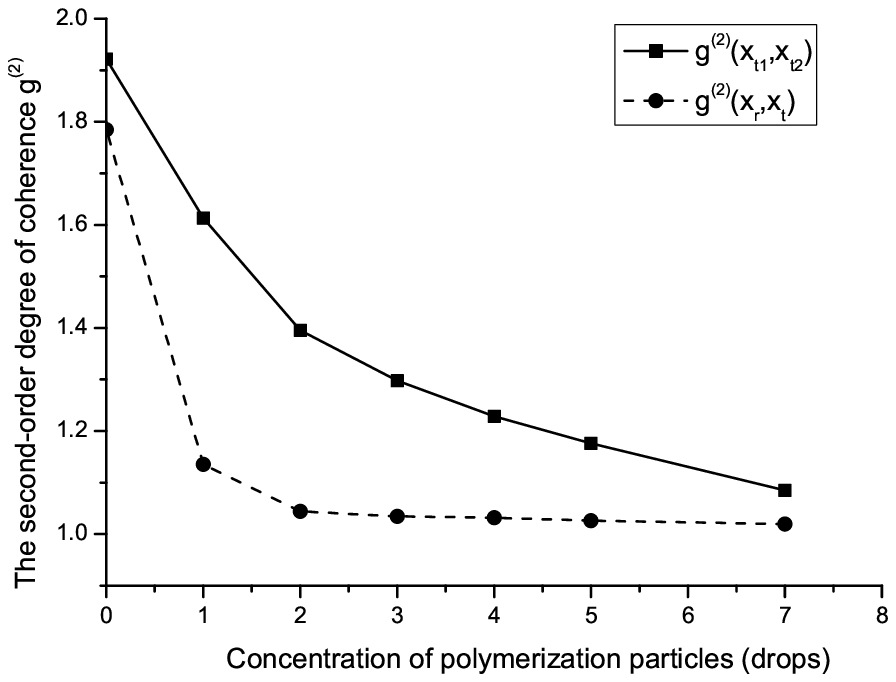}}
\end{figure}

\begin{figure}
\centerline{
\includegraphics[width=8cm]{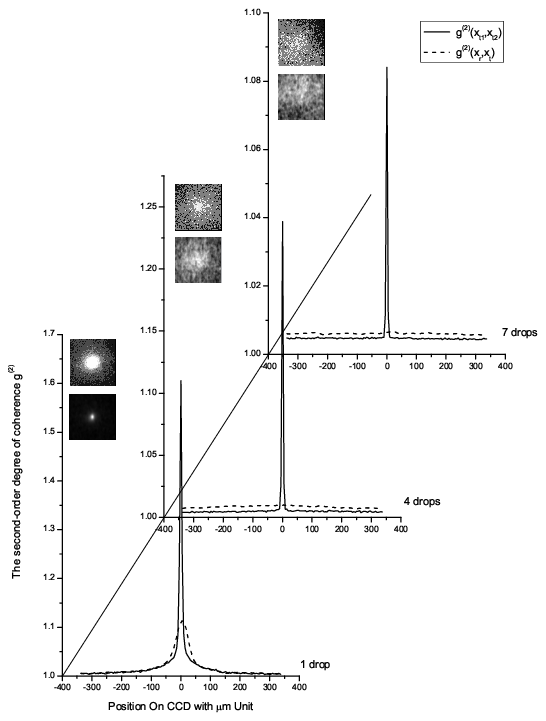}}
\caption{The effect of concentration of emulsion polymerization
particles on $g^{(2)}$ when $L_1$=40mm and $L_2$=0mm without the
object.}
\end{figure}

\begin{figure}
\centerline{
\includegraphics[width=8cm]{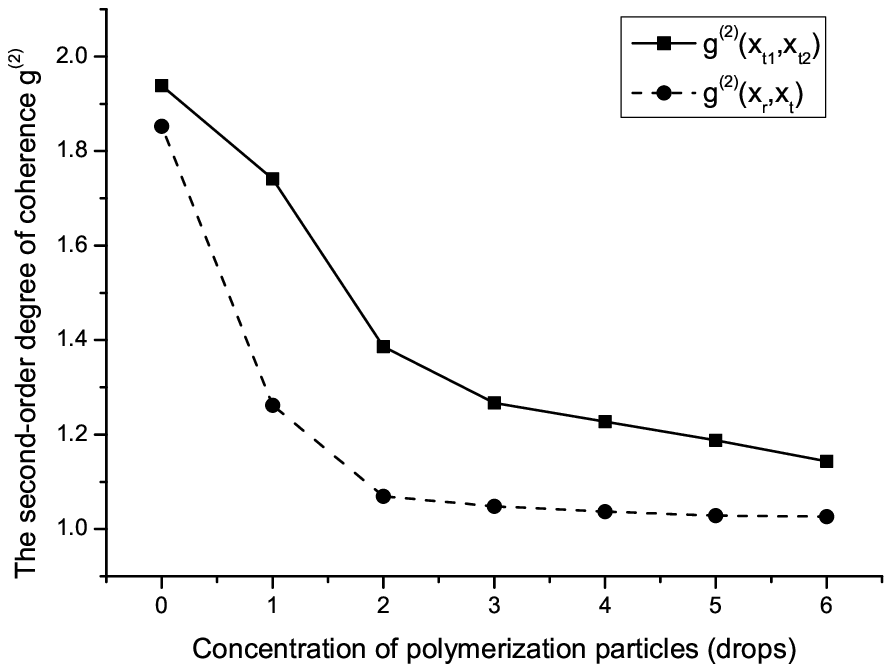}}
\end{figure}

\begin{figure}
\centerline{
\includegraphics[width=8cm]{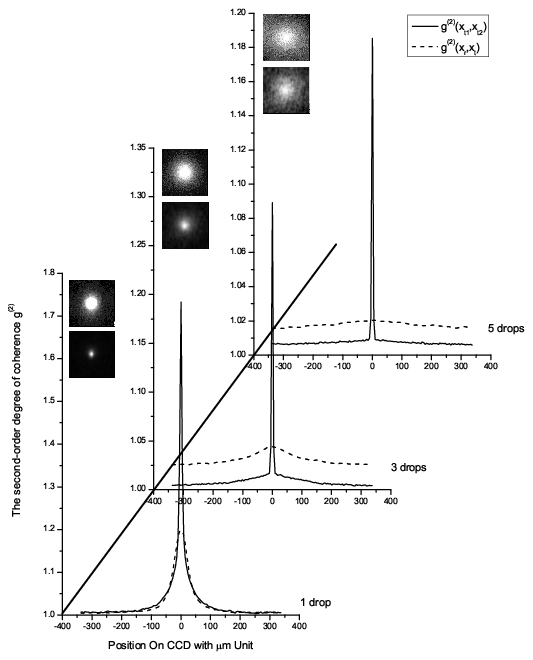}}
\caption{The effect of concentration of emulsion polymerization
particles on $g^{(2)}$ when $L_1$=20mm and $L_2$=20mm without the
object.}
\end{figure}

When there are five drops emulsion polymerization particles put into
the solution $NaCl$, the intensity distribution on the detector
$D_r$ is heterogenous because there is no multiple scattering.
However, we can find the homogeneous intensity distribution on the
detector $D_t$ (in Fig. 12).

\begin{figure}
\centerline{
\includegraphics[width=8cm]{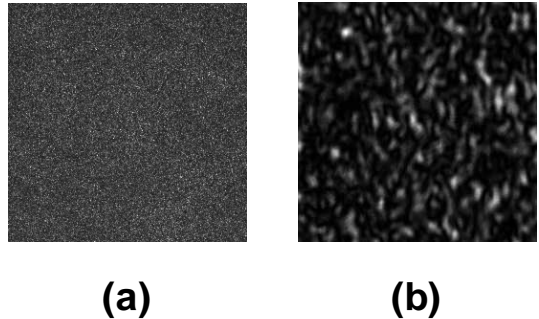}}
\caption{Images from a single frame speckle distribution when
$L_1$=20mm and $L_2$=20mm without the object and there are five
drops emulsion polymerization particles put into the solution
$NaCl$, (a). from the CCD camera $D_t$, (b). from the CCD camera
$D_r$.}
\end{figure}

\section{discussion and inclusion}
When the object is illuminated directly by thermal light, multiple
scattering between the object plane and the detector $D_t$ is the
main reason leading to the degradation of the qualities of images
(Fig. 3). In scattering media, different from images obtained by the
first-order correlation of light field, ghost imaging causes the
separation of detection and imaging. Even if the information from
the object is distorted by multiple scattering, there is no effect
on ghost imaging. Because multiple scattering destroys the
correlation of the light field, the visibility of ghost images is
degraded but the resolutions doesn't degrade. However, the
visibility of ghost images can be enhanced by means of the decrease
of the resolution \cite{Shen,Cai1,Cai2,Gatti4}. The experimental
result in Fig. 9 (c) also demonstrated that. Transmission
information of the light on the object plane (namely $\alpha_1$) has
a great effect on the visibility of ghost images. Because multiple
scattering leads to the decay of the correlation between the test
and reference pathes with the decrease of Transmission information
of the light (Figs 8 and 9). The higher the Transmission information
is, the better visibility the ghost image has. When there is no
multiple scattering between the object plane and the source, then
ballistic component of the light on the object plane is equal to 1,
so we can always obtain a ghost image with high quality whether
there is multiple scattering between the object plane and the
detector $D_t$. Otherwise, because the ballistic component of the
light is impossible to be 0, we can always see the shapes of the
object (Figs. 3(b) and 6). The degradation of the qualities of the
images is provoked by the decay of visibility in scattering media,
which is different absolutely from the low resolution caused by
diffraction-limited. When the object is fixed in the middle of the
scattering media, as shown in Fig. 11, even if the second-order
degree of coherence $g^{(2)}$ is very low, we can obtain a ghost
image with much better quality than the image gained when the object
is illuminated directly by thermal light. By the characteristics of
ghost imaging, the qualities of ghost images does not depend on the
intensity of the light but on the intensity fluctuations. As the
multiple scattering increased, the intensity fluctuations will
reduce (Figs 10 and 11), which leads to the degradation of the
visibility of the image. The expression in the Eq. (15) can explain
basically the experimental results and the fact discussed above,
which shows that the results described by the Eq. (15) are
reasonable for the description of ghost imaging in scattering media.
However, when the test detector is a bucket detector, the qualities
of ghost images depend on the multiple scattering between the object
plane and the source, which is opposite to imaging by the direct
illumination of the thermal light (Fig. 3(a)). By the results in
Fig. 8 (a) and (c), maybe we can also obtain a ghost image with high
visibility by means of eliminating the multiple scattering between
the object plane and the source or between the object plane and the
detector $D_t$. One is like the technique of DOT. Firstly we obtain
a improved image by iterative recovery method of DOT from a single
frame of image collected by the detector $D_t$, then we will get a
ghost image with improved visibility by the correlation measurement
of intensity fluctuations between the detector $D_r$ and the
improved images. The other is that we get the intensity distribution
on the object plane by the measurement to transmission and
reflection coefficients, then do the correlation measurement of the
intensity fluctuations between the detector $D_t$ and intensity
distribution obtained by numerical simulation. The method discussed
in Fig. 7 can improve the visibility of ghost images to some extent.
Generally speaking, by ghost imaging method, we can always obtain a
image with much better quality on the base of the image obtained by
a novel conventional imaging technique with thermal light. For
entangled source, because of the entanglement characteristic of the
two photons, maybe we can also improve the visibility of ghost
images. Based on the effect of photon bunching, and the new source
with high $g^{(2)}$ can be obtained, the visibility of ghost images
may also be enhanced obviously.

In medical science, in order to avoid ionizing radiation of X-ray,
optical photons provides nonionizing and safe radiation for medical
applications. Recently there has been increasing interest in the
field of the imaging, test and diagnosis of the biological tissues
with the infrared and the near infrared light
\cite{Wang,Sachin,Brian}. Because the near infrared light around
700-nm wavelength can penetrate several centimeters into biological
tissue \cite{Brian}. But several factors still limit the imaging
quality. Because most biological tissues are characterized by strong
optical scattering and hence are referred to as scattering media or
turbid media. The images suffer reduced resolution and contrast due
to multiple scattering, which leads to a low efficiency and accuracy
of diagnosis and a difficulty of analysis in medical science. So the
diffusion-like behavior of light in biological tissue presents a key
challenge for optical imaging. The ghost imaging discussed here may
solve the problems about the low quality of the imaging in
biological tissues.

In conclusion, for the first time, detection and imaging of the
object information are separated by ghost imaging, which provides a
new way for imaging in scattering media. we have demonstrated
experimentally and theoretically for the first time that the images
with high qualities in a scattering media can still be obtained for
ghost imaging when the test detector is an array of pixel detector
and there is only multiple scattering between the object plane and
the source or between the object plane and the detector. When the
object is fixed in the middle of scattering media, we can also gain
ghost images with much better qualities than the images obtained
when the object is illuminated directly by the thermal light. The
probability amplitudes of the incident light $\alpha_1$ and
$\alpha_2$ have a great effect on the visibility of ghost images.
These results will be very useful for the imaging and diagnosis in
medical science.

The work was partly supported by the Hi-Tech Research and
Development Program of China, Project No. 2006AA12Z115, and Shanghai
Fundamental Research Project, Project No. 06JC14069.

\end{document}